\documentclass[preprint,notoc]{JHEP3}
\usepackage{epsfig}
\def\eslt{\not\!\!{E_T}}
\def\to{\rightarrow}

\def\bi{\begin{itemize}}
\def\ei{\end{itemize}}

\def\tst{\tilde t}
\def\ttau{\tilde \tau}

\def\tg{\tilde g}

\def\tell{\tilde\ell}

\def\tw{\widetilde W}
\def\tz{\widetilde Z}

\def\agt{\stackrel{>}{\sim}}

\title{
Reach of the Fermilab Tevatron for minimal
supergravity in the region of large scalar masses
}

\author{Howard Baer and Tadas Krupovnickas
\\ Department of Physics, Florida State University\\ 
Tallahassee, FL 32306, USA\\
E-mail: \email{baer@hep.fsu.edu}, \email{tadas@hep.fsu.edu}}
\author{Xerxes Tata
\\ Department of Physics and Astronomy, University of Hawaii,\\
Honolulu, HI 96822, USA \\ 
E-mail: \email{tata@phys.hawaii.edu}}

\preprint{\vbox{\hbox{FSU-HEP-030530} \vspace{0.2cm}
                \hbox{UH-511-1028-03}}} 

\abstract{
The reach of the Fermilab Tevatron for supersymmetric matter has been
calculated in the framework of the minimal supergravity model
in the trilepton channel. Previous analyses of this 
channel were restricted to scalar masses 
$m_0\le 1$ TeV. We extend the analysis
to large values of scalar masses $m_0\sim 3.5$ TeV, in order to probe
the compelling hyperbolic branch/focus point (HB/FP) region,
where the superpotential $\mu$ parameter becomes small, and 
which is one of the mSUGRA parameter space
regions consistent with WMAP data. In this region,
assuming a $5\sigma$ ($3\sigma$) signal with 10 (25) fb$^{-1}$ 
of integrated luminosity, the Tevatron reach in the 
trilepton channel extends up to
$m_{1/2}\sim 190$ (270) GeV independent of $\tan\beta$.  
This corresponds to a reach in terms of the gluino mass of 
$m_{\tg}\sim 575$ (750) GeV. 
}


\begin{document}

\section{Introduction}
\label{sec:intro}

Run 2 of the Fermilab Tevatron $p\bar{p}$ collider has begun
at center of mass energy $\sqrt{s}=1.96$ TeV, and already
the CDF and D0 experiments have gathered over 100 pb$^{-1}$ 
of integrated luminosity. Projections are to acquire anywhere from
2-25 fb$^{-1}$ of integrated luminosity before turn on of the
CERN LHC. One prominent goal of Tevatron experiments is to
discover the Higgs boson, which may well be within reach according to 
analyses of electroweak radiative corrections. Another prominent goal
is to obtain evidence for weak scale supersymmetric matter.

The search for supersymetry is somewhat model dependent. 
In this paper, we adopt the paradigm
minimal supergravity model (mSUGRA)\cite{msugra},
with parameters
\begin{equation}
m_0,\ m_{1/2},\ A_0,\ \tan\beta ,\ sign(\mu ) .
\end{equation}
In 
models such as mSUGRA, with gaugino mass unification and a 
weak scale gravitino mass, the gluino to chargino mass ratio is 
$m_{\tg}/m_{\tw_1}\sim 3.7$, so that
bounds from LEP2 ($m_{\tw_1}>103.5$ GeV)\cite{lep2} likely place
gluino pair production out of reach of Tevatron experiments.
Since squark masses are usually comparable to or greater
than $m_{\tg}$, it is likely that squark pair production is
beyond the Tevatron reach as well. An exception occurs for
third generation squarks- the top and bottom squarks- since these
might have much lower masses\cite{stops}. 
In addition, slepton pair production occurs at
low enough rates in these models that they are unlikely to be 
directly observable\cite{sleptons}.

However, charginos and neutralinos may well be within the kinematic
reach of the Tevatron, and can be produced with observable cross sections.
Importnat pair production reactions include
\begin{itemize}
\item $p\bar{p}\to \tw_1\tz_1 X$, 
\item $p\bar{p}\to\tw_1^+\tw_1^- X$ and 
\item $p\bar{p}\to\tw_1\tz_2 X$,
\end{itemize}
where $X$ represents assorted hadronic debris. The purely hadronic
final states suffer large QCD backgrounds, while the leptonic final
states have more manageable electroweak backgrounds. The first of these
reactions can lead to single lepton plus missing energy states, which
suffer large backgrounds from $W\to \ell\nu_\ell$ 
production (here, $\ell = e,\ \mu$ and $\tau$). The second
chargino pair reaction suffers large backgrounds from $WW$ and 
$Z\to\tau\bar{\tau}$ production. The last of these--
$\tw_1\tz_2$ production-- can lead
to clean (non-jetty) trilepton plus $\eslt$ final states which can
be above SM background levels for significant regions of model parameter
space.

The clean trilepton signature was suggested as long ago as 1983\cite{dicus},
and explicit collider calculations for production via 
on-shell gauge bosons were performed in Refs.\cite{bt1,bht},
including spin correlations between initial and final states.
Arnowitt and Nath pointed out that rates may be detectable even for 
production via off-shell gauge bosons\cite{an}. More detailed
projections (based on partial neutralino branching fraction
calculations) yielded a pessimistic assessment of the Tevatron reach\cite{bm}.
Improved sparticle production and decay calculations however 
showed that in fact
the reach of Fermilab Tevatron experiments could extend well past
LEP2 for significant regions of model parameter space\cite{bt}.
This was followed by a number of calculations\cite{dimitri} 
and collider simulations of 
clean trilepton detection rates  considered against SM backgrounds
arising mainly from $WZ$ production\cite{bkt,kane,tevreach,mp1},
with results being extended to large $\tan\beta$ in Ref. \cite{ltanb}.
Especially for large $\tan\beta$, it was found that the greatest reach
was obtained via the inclusive trilepton channel, with jetty events
allowed into the trilepton sample\cite{bdpqt,bkao}. 
At the Fermilab Tevatron SUSY/Higgs workshop (concluded in year 
2000), it was found that in fact the largest backgrounds came from
off-shell $W^*Z^*$ and $W^*\gamma^*$ production\cite{campbell}. These backgrounds were
calculated, and cuts were modified to show that in fact the inclusive trilepton
signal was still observable over large portions of mSUGRA model
parameter space\cite{bdpqt,bkao,mp2,sugra_report,dedes}. 
Reach calculations were made in the $m_0\ vs.\ m_{1/2}$ plane 
extending out to $m_0$ values as high as 1 TeV.

Since these previous calculations, a greater emphasis has been placed
on mSUGRA model parameter space at large $m_0$ values. It has been
noticed that as $m_0$ increases, ultimately the superpotential
$\mu$ parameter, as derived from radiative electroweak symmetry
breaking (REWSB), becomes small in magnitude shortly before
encountering the region where REWSB breaks down\cite{bcpt}. 
Chan {\it et al.}\cite{ccn},
adopting effectively $\mu^2/M_Z^2$ as a fine-tuning parameter,
emphasized that the entire region of small $\mu^2$ at 
large $m_0$ may be considered to 
have low fine tuning; they dubbed the region as the ``hyperbolic branch''. 
Later, using more sophisticated fine tuning calculations, 
Feng {\it et al.}\cite{fmm}
showed that just the low $m_{1/2}$ portion of the hyperbolic branch
has low fine tuning. The peculiar focussing behavior of the RG running 
of the soft breaking
Higgs mass $m_{H_u}^2$ in this region led to the characterization 
as the ``focus point'' region. 
In this paper, we will refer to the
large $m_0$ region with small $|\mu |$ as the hyperbolic branch/focus point
(HB/FP) region.

The large $m_0$ region of parameter space has received renewed attention
as well due to several experimental developments. First, 
improved evaluations of the neutralino relic 
density\cite{ellis_co,Afunnel,fmw,st_co,relic} 
show four viable 
regions of mSUGRA model parameter space consistent with recent WMAP and 
other data sets\cite{wmap}. 
These include 1.) the bulk region at low $m_0$ and $m_{1/2}$
where neutralinos may annihilate in the early universe via $t$-channel
slepton exchange, 2.) the stau co-annihilation region where 
$m_{\tz_1}\simeq m_{\ttau_1}$\cite{ellis_co}, 
3.) the axial Higgs $A$ annihilation
corridor at large $\tan\beta$\cite{Afunnel} and 4.) 
the HB/FP region where the neutralino
has a significant higgsino component and can readily annihilate to
$WW$ and $ZZ$ pairs in the early universe\cite{fmw}. 
A fifth region of 
squark-neutralino co-annihilation can exist as well for particular
values of the $A_0$ parameter that give rise, for instance, to
$m_{\tst_1}\simeq m_{\tz_1}$\cite{st_co}. 

The bulk region of relic density, which originally seemed most compelling,
is difficult to reconcile with LEP2 limits on the Higgs mass, the
$b\to s\gamma$ branching fraction, and for $\mu <0$, the muon
anomalous magnetic moment\cite{constr,sug_chi2}. 
The stau co-annihilation region is viable,
but 
unless the parameters are just right, the relic density can
become either too large or too small.
The $A$-annihilation corridor is also viable, but requires large
$\tan\beta$, and usually sparticle masses are beyond the reach of
Tevatron searches. The HB/FP region remains viable for almost all $\tan\beta$
values, and since scalar sparticles are typically in the multi-TeV regime,
gives a value of $b\to s\gamma$ and $a_\mu$ in close accord with SM
predictions\cite{constr,sug_chi2}. 
Since $|\mu |$ is small in the HB/FP region, then charginos and 
neutralinos are expected to be light, and hence signals such as the
trilepton one may be accessible to Tevatron collider searches.

For these reasons, in this paper we extend the trilepton search results 
presented in Ref. \cite{bdpqt} to large values of $m_0>1$ TeV, including the
HB/FP region. For our signal calculations, we use Isajet v7.66\cite{isajet}.
This version of Isajet contains 1-loop corrections to all sparticle 
masses\cite{pierce}, and treats the Higgs potential in the RG-improved one loop
effective potential approximation. It yields good overall agreement
with other publicly available codes, as documented by
Allanach {\it et al.}\cite{kraml}, including the location of the HB/FP region.
We note, however, that the location of the HB/FP region is very sensitive 
to the value of $m_t$ adopted in the calculation.
To illustrate this, we show in Fig. \ref{fig:10ptop} the boundary of 
parameter space in the $m_0\ vs.\ m_{1/2}$ plane for $A_0=0$, $\tan\beta =10$,
$\mu >0$, and for $m_t=172.5,\ 175,\ 177.5$ and 180 GeV. The right-hand 
boundary, which dictates the location of the HB/FP region,
ranges from 2-20 TeV depending on $m_t$ and $m_{1/2}$.
\FIGURE[t]{\epsfig{file=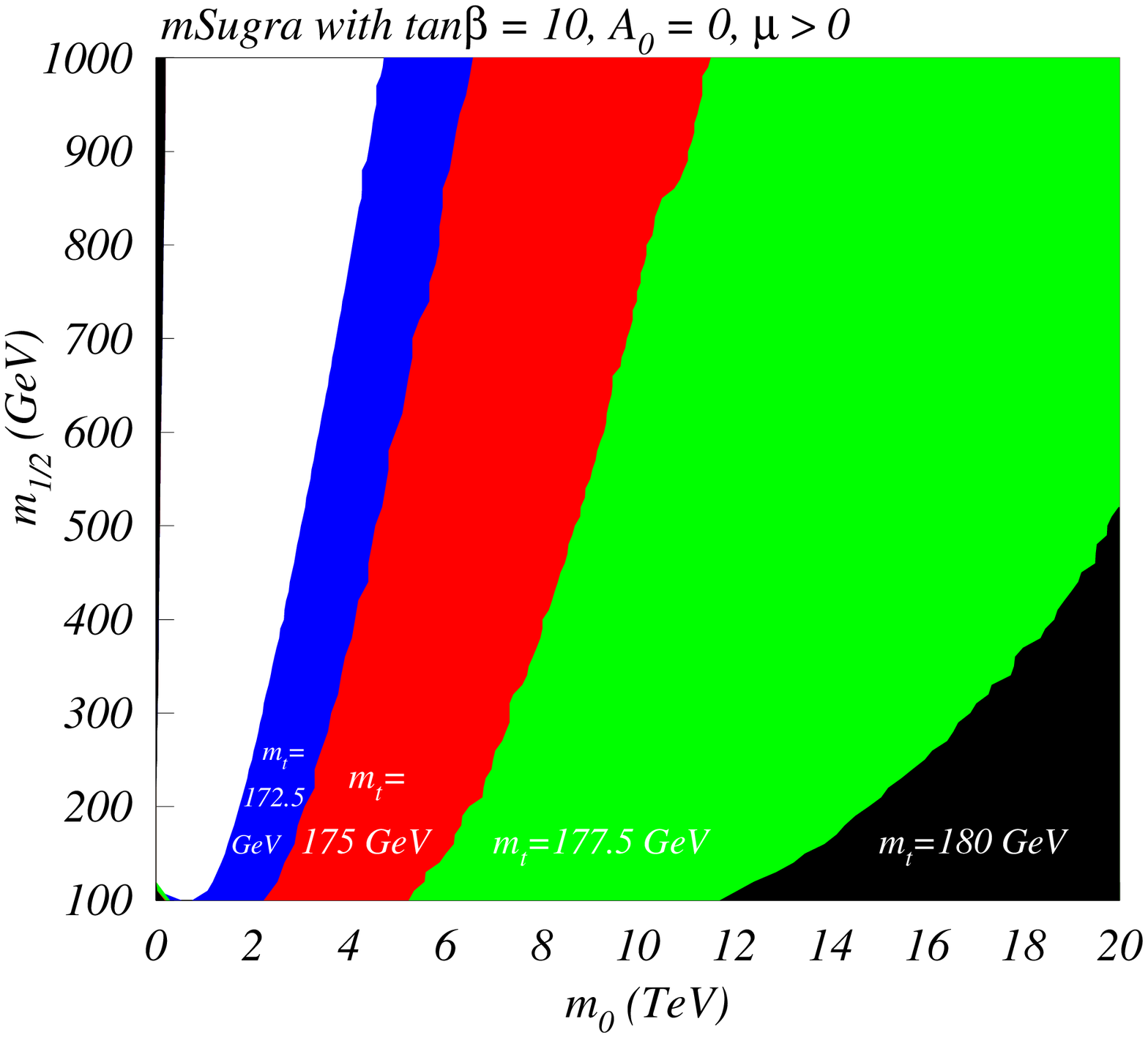,width=15cm} 
\caption{Boundary of the 
$m_0\ vs.\ m_{1/2}$ parameter plane 
of the mSUGRA model, with
$\tan\beta =10$, $A_0=0$ and $\mu >0$,
for $m_t=172.5,\ 175,\ 177.5$ and 180 GeV.
}
\label{fig:10ptop}}

We adopt the SM background calculation as presented in Ref. \cite{bdpqt}.
The backgrounds evaluated include $WZ\ (Z\to\tau\bar{\tau})$
production, $Z^*Z^*$ production, $t\bar{t}$ production 
and trilepton production through
a variety of $2\to 4$ Feynman graphs including $W^*\gamma^*$
and $W^*Z^*$ production, as calculated using Madgraph\cite{madgraph}.
In Ref. \cite{bdpqt}, a variety of cuts were proposed to reduce
background compared to signal. Here, we adopt set SC2 from Ref. \cite{bdpqt},
which generally gave a reach in accord with calculations from
Refs. \cite{bkao,mp2}. For these cuts, the total $3\ell +\eslt$
background level was found to be 1.05 fb.

Our first results are shown in Fig. \ref{fig:10p}, 
where we show the $m_0\ vs.\ m_{1/2}$ plane for $\tan\beta =10$,
$A_0=0$ and $\mu >0$. Here, and in the rest of this
paper, we fix $m_t=175$~GeV. The red regions are excluded
by lack of REWSB (right side) and presence of a stau LSP (left side).
The magenta shaded regions are excluded by the LEP2 bound
$m_{\tw_1}>103.5$ GeV.\footnote{LEP experiments
exclude charginos up to 91.9~GeV even if $m_{\tw_1}-m_{\tz_1}$ is as
small as 3~GeV, so that the LEP excluded region is unlikely to be much
altered even in the FP/HB part of parameter space.} In addition, the region below the magenta contour
has $m_h<114.1$ GeV, in violation of LEP2 limits on the search for
a SM Higgs boson. We also show the Tevatron reach contours
requiring a $5\sigma$ signal for 10 fb$^{-1}$ of integrated 
luminosity (solid contour),
and a more optimistic contour for a $3\sigma$ signal for 
25 fb$^{-1}$ (dashed contour).
These correspond to signal cross sections rates of 1.62 fb and
0.61 fb, respectively, after application of cuts SC2 of Ref.~\cite{bdpqt}.

The first feature to note is that the LEP2 bound on $m_h$ 
now excludes essentially all the region that was previously mapped
out in Refs.~\cite{bdpqt,bkao,mp2}. There is some uncertainty
of a few GeV with respect to the calculation of 
$m_h$ (see {\it e.g.} Ref.~\cite{kraml}), so the magenta contour is
not a solid bound on mSUGRA parameter space. 
In any case, the reach region of the Fermilab Tevatron separates into
two regions. The first, for very low $m_0$ values where sleptons are light,
has the $\tz_2\to\tell\ell$ and $\tw_1\to\tell\nu$ two body 
decay modes allowed, 
which dominate the $\tz_2$ and $\tw_1$
branching fractions. The large leptonic branching fractions 
give rise to high rates for  trileptons.
The second region occurs for $m_0\agt 300$ GeV. 
As $m_0$ increases, the slepton masses also increase so that
two-body chargino and neutralino decay modes become forbidden.
In the region of moderate $m_0\sim 200$ GeV, three-body
decays such as $\tz_2\to\ell\bar{\ell}\tz_1$ can occur, but
interference between slepton- and $Z$-mediated decay graphs
give rise\cite{bt} to a very tiny leptonic branching fraction for
the $\tz_2$, and hence a sharp drop in the Tevatron
reach for SUSY via trileptons. As $m_0$ increases further, the slepton
mediated decay diagrams for $\tz_2$ three-body decay are 
increasingly suppressed, and the decay rate becomes dominated by the 
$Z$ exchange graph. Ultimately, the branching fraction
$\tz_2\to e^+e^-\tz_1$ increases to $\sim 3\%$, {\it i.e.} the same as
the $Z$ branching fraction to electrons. Thus the reach of 
Fermilab Tevatron experiments increases and levels off as 
$m_0$ becomes large. However, for very large $m_0$ values, 
then we enter the HB/FP region, where $|\mu|$ become small. In this case,
chargino and neutralino masses decrease, and the production cross sections
rise, yielding an increased reach at very large $m_0$. 
From Fig.~\ref{fig:10p}, we see that the $5\sigma$ reach for 10 fb$^{-1}$
reaches $m_{1/2}\sim 175$ GeV for $m_0\sim 1000-2000$ GeV, corresponding
to a reach in $m_{\tw_1}$ ($m_{\tg}$) of 125 (525) GeV.
The observability of the $3\sigma$ signal for 25 fb$^{-1}$ of
integrated luminosity extends to values of $m_{1/2}\sim 210$ GeV,
corresponding to values of $m_{\tw_1}$ ($m_{\tg}$) $\sim 150$ (600) GeV.  
In the HB/FP region, this extends to
$m_{1/2}\sim 270$ GeV, corresponding to a reach in 
$m_{\tg}\sim 750$ GeV.
\FIGURE[t]{\epsfig{file=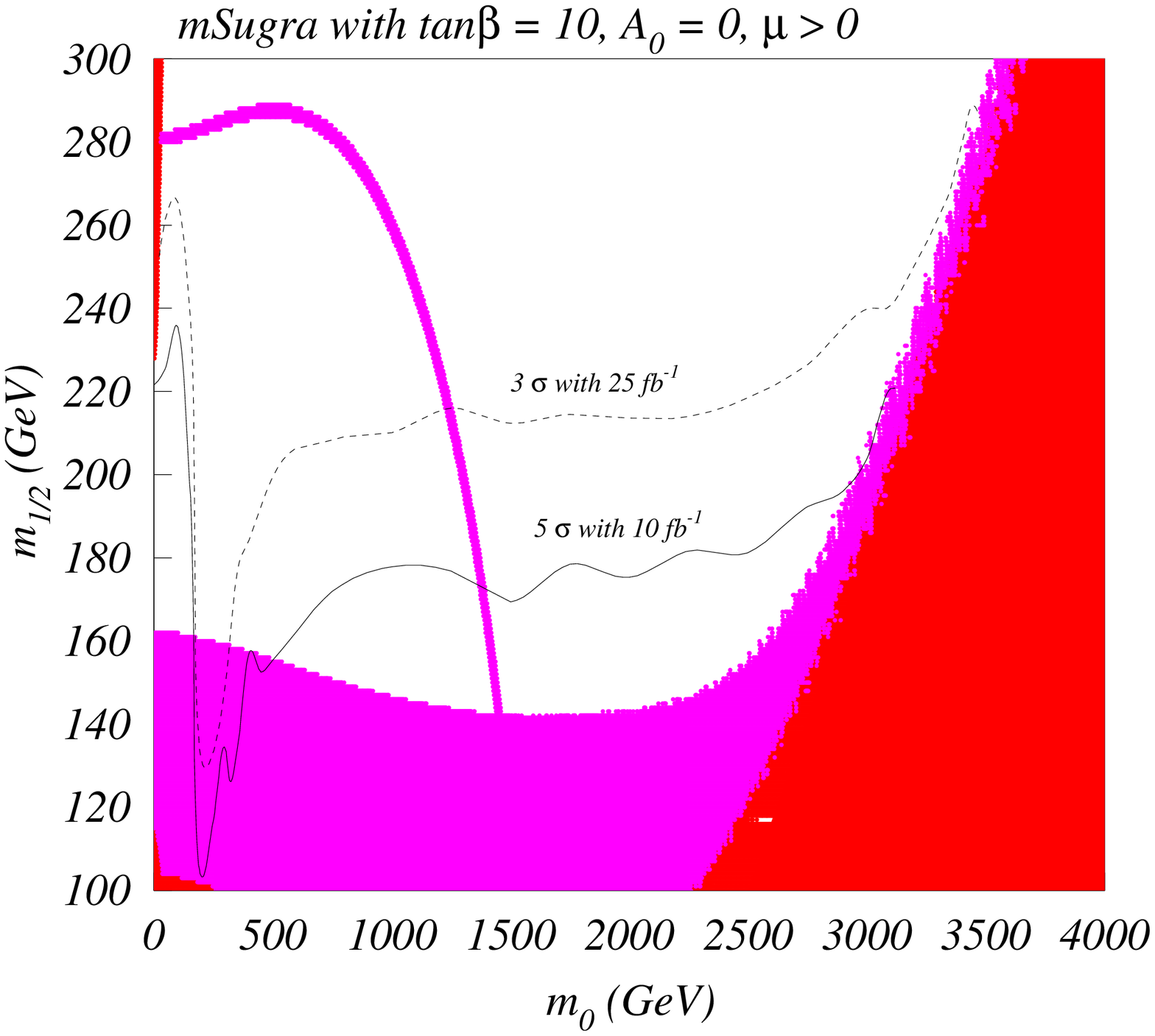,width=15cm} 
\caption{The reach of Fermilab Tevatron in the 
$m_0\ vs.\ m_{1/2}$ parameter plane 
of the mSUGRA model, with
$\tan\beta =10$, $A_0=0$ and $\mu >0$,
assuming a $5\sigma$ signal at 10 fb$^{-1}$ (solid) and a $3\sigma$ 
signal with 25 fb$^{-1}$ of integrated luminosity (dashed). The red 
(magenta) region is
excluded by theoretical (experimental) constraints. The region below
the magenta contour has $m_h <114.1$ GeV, in violation of Higgs mass
limits from LEP2. 
}
\label{fig:10p}}

To gain a better understanding of what's happening in the HB/FP region, 
in Fig.~\ref{fig:mcs}{\it a.}) we plot the masses of various charginos and
neutralinos and the $\mu$ parameter as a function of $m_0$
for fixed $m_{1/2}=225$ GeV, $A_0=0$, $\tan\beta =30$ and $\mu >0$.
Initially, for $m_0\sim 1500-1700$ GeV, the $\tw_1$, $\tz_1$ and
$\tz_2$ masses are essentially constant
with $m_0$, as might be expected. As we approach the large $m_0$
HB/FP region, the value of $\mu$ drops, and consequently the light
chargino and neutralino masses drop, as they become increasingly
higgsino-like. As $\mu \to 0$, $m_{\tw_1}-m_{\tz_1}$ also approaches
zero. However, the LEP2 limit of $m_{\tw_1}=103.5$ is reached before
the $\tw_1$ and $\tz_1$ become nearly degenerate.

\FIGURE[t]{\epsfig{file=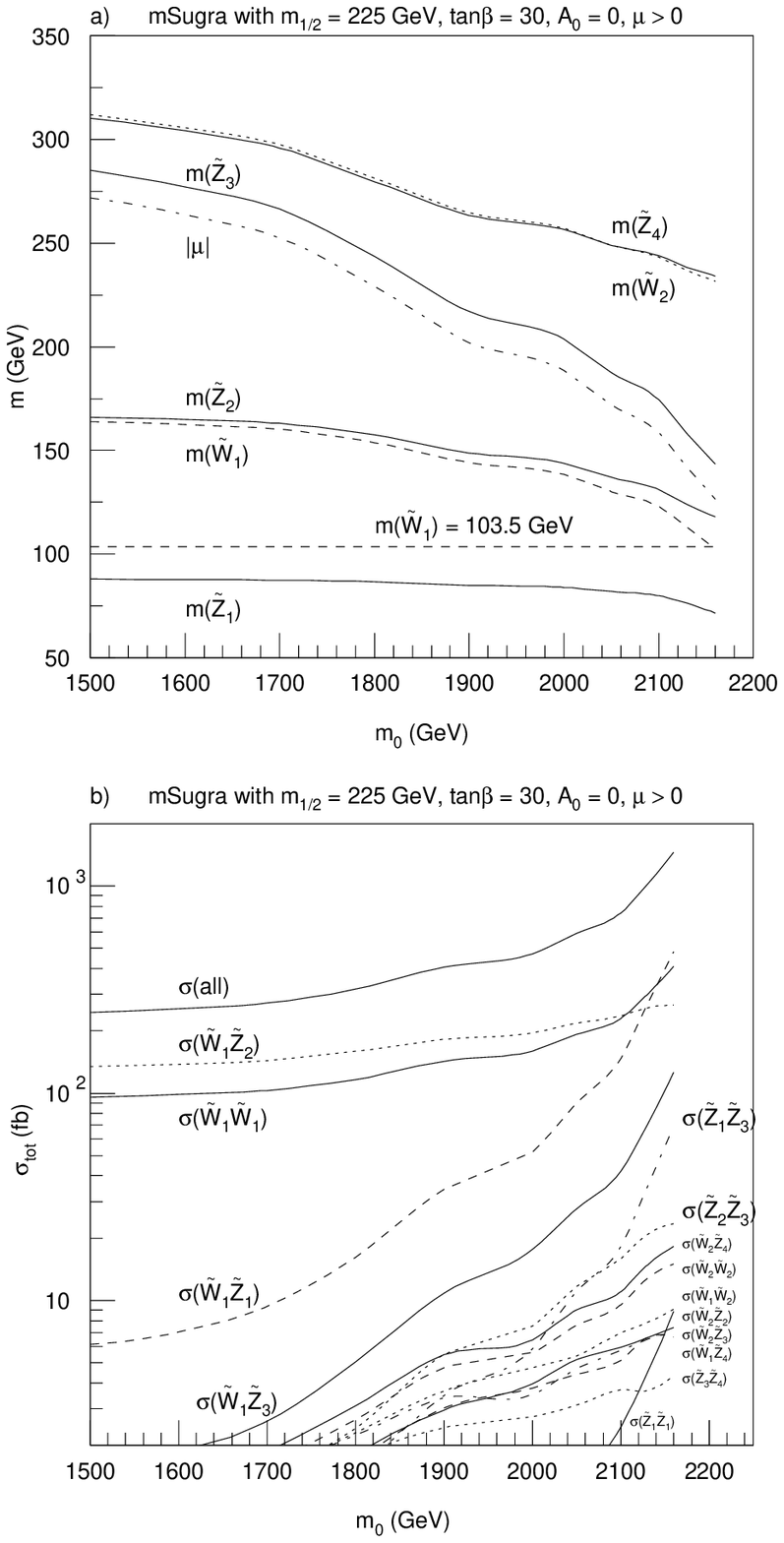,width=15cm} 
\caption{In {\it a.}), we show selected sparticle masses versus $m_0$
in the HB/FP region. In {\it b.}), the corresponding total cross sections
are shown. 
}
\label{fig:mcs}}

In Fig.~\ref{fig:mcs}{\it b}.), we also show various chargino and neutralino
cross sections versus $m_0$ for the same parameters as
in Fig.~\ref{fig:mcs}{\it a}.). 
For intermediate values of $m_0$, $\sigma (\tw_1\tz_2)$ and 
$\sigma (\tw_1^+\tw_1^-)$
are dominant.
As one increases $m_0$ and approaches the HB/FP region, the various
chargino and neutralino masses drop, and the production cross sections
increase, giving rise to an increased reach by Tevatron experiments.
At the highest $m_0$ values, actually $\sigma (\tw_1\tz_1)$ production has
become dominant. In addition, a variety of cross sections such
as $\sigma (\tz_1\tz_3)$, $\sigma (\tz_2\tz_3)$, $\sigma (\tw_2\tz_4)$, 
$\cdots$ are
increasing, and their sum can be non-negligible. These heavier -ino
states in general have lengthier cascade decays, and can lead to
complicated signals including multileptons which may be 
at the edge of observability.

Fig.~\ref{fig:30p} shows the Tevatron reach 
in the $m_0\ vs.\ m_{1/2}$ plane for $\tan\beta =30$, $A_0=0$
and $\mu >0$. In this case, the reach at large $m_0$ remains
large as in the $\tan\beta =10$ case from Fig.~\ref{fig:10p}.
However, the reach at low $m_0$ has diminished somewhat,
which is an effect of large $\tan\beta$ where the $\tau$ and
$b$ Yukawa couplings become large, and the $\ttau_1$ mass
becomes lighter than that of other sleptons. 
The enhanced chargino and neutralino decays to taus in this region
comes at the expense of decays to $e$s and $\mu$s, so that the
low $m_0$ $3\ell$ reach is diminished\cite{ltanb}. 

\FIGURE[t]{\epsfig{file=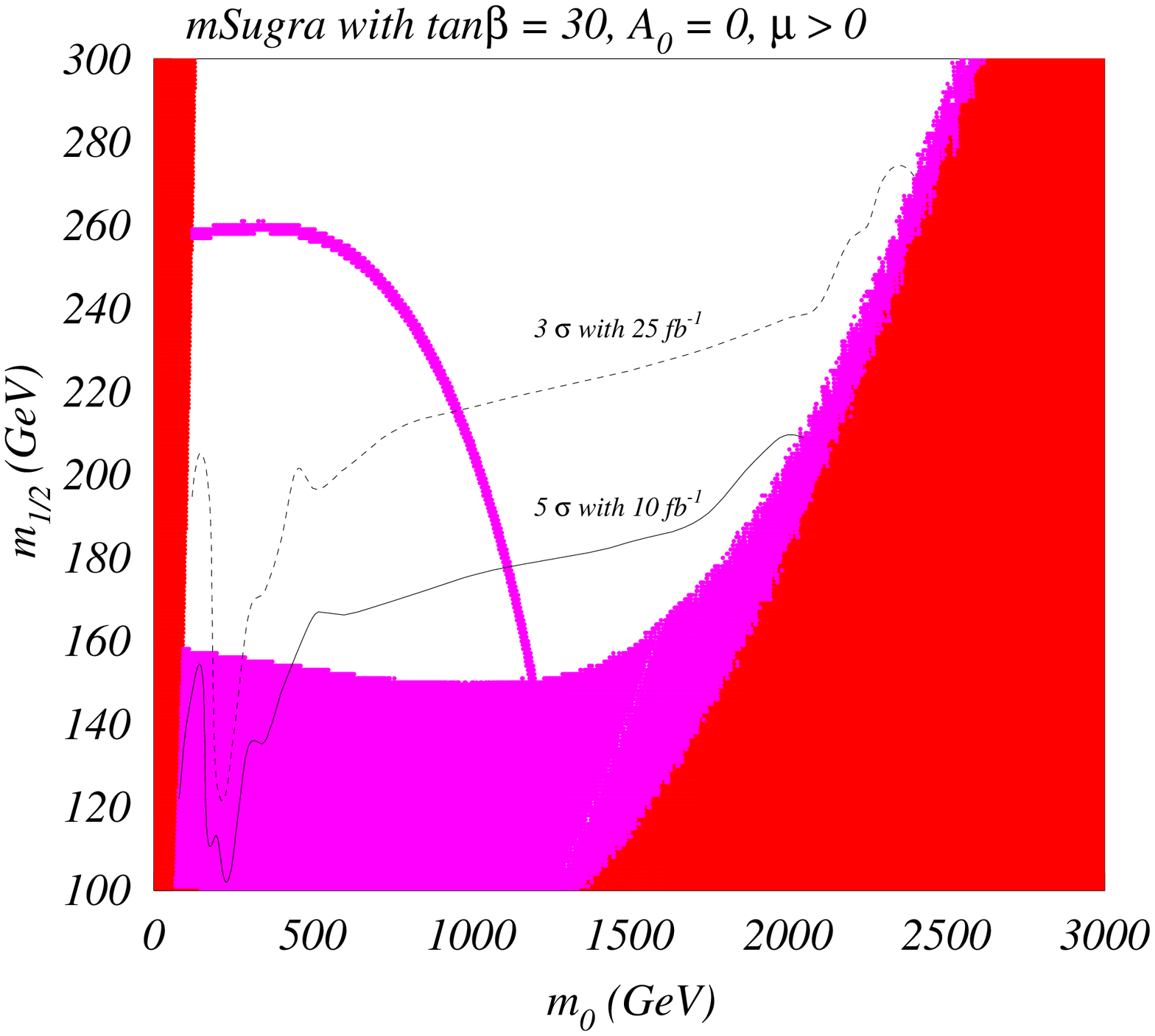,width=15cm} 
\caption{The reach of Fermilab Tevatron in the 
$m_0\ vs.\ m_{1/2}$ parameter plane 
of the mSUGRA model, with
$\tan\beta =30$, $A_0=0$ and $\mu >0$.
The red (magenta) region is
excluded by theoretical (experimental) constraints. The region below
the magenta contour has $m_h <114.1$ GeV, in violation of Higgs mass
limits from LEP2.
}
\label{fig:30p}}

The $m_0\ vs.\ m_{1/2}$ plane is shown for $\tan\beta =45$, 
with $\mu <0$ in Fig.~\ref{fig:45m}. The plot shows even greater
reach suppression at low $m_0$ due to the increase in $\tan\beta$,
where an even greater suppression of chargino and neutralino decays to
$e$s and $\mu$s occurs at the expense of decays to $\tau$s. Irregardless,
as $m_0$ increases, sleptons, smuons and staus all decouple from the
decay calculations, so that results are relatively insensitive to
$\tan\beta$, and the reach remains large in the HB/FP region.

\FIGURE[t]{\epsfig{file=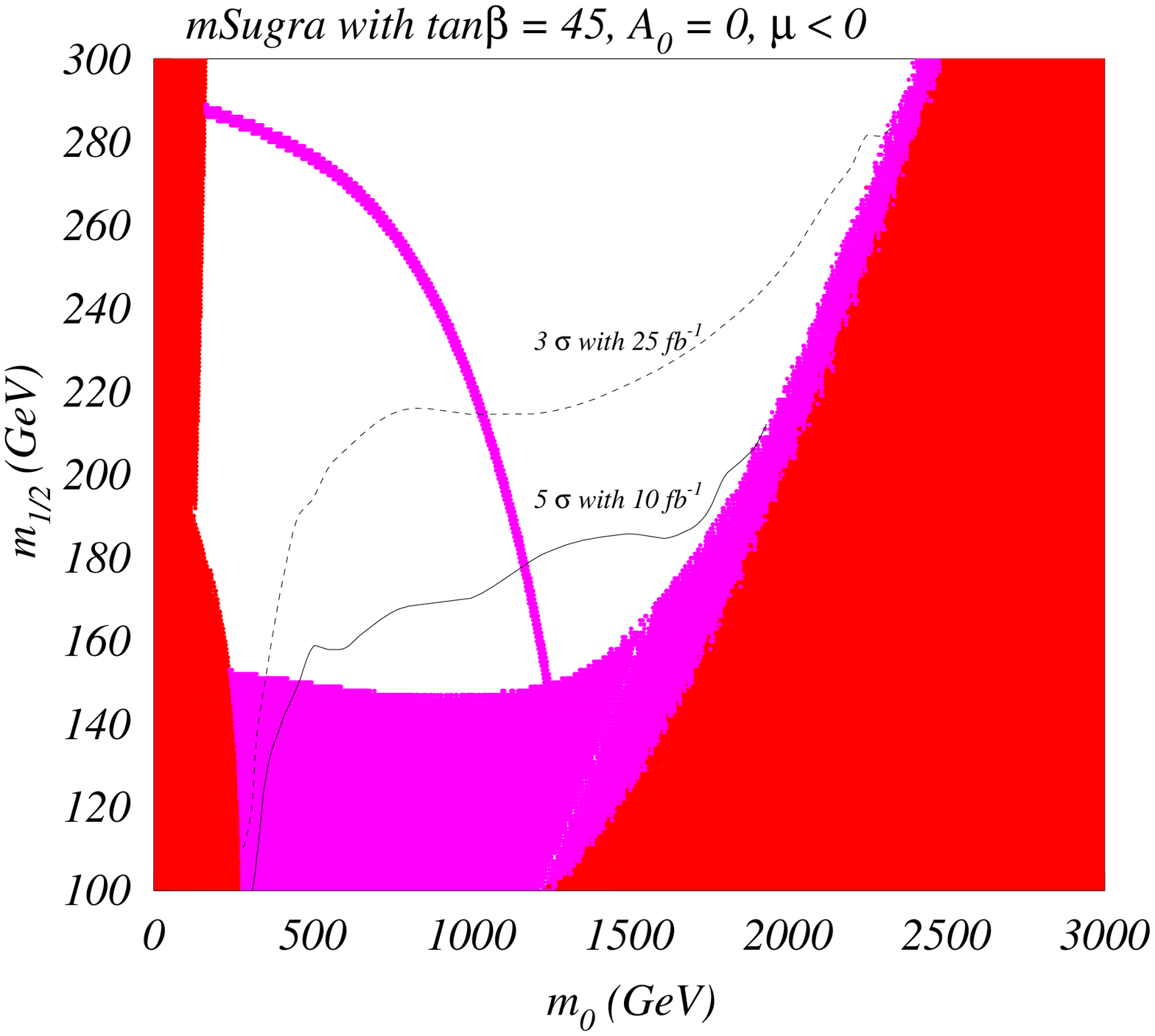,width=15cm} 
\caption{The reach of Fermilab Tevatron in the 
$m_0\ vs.\ m_{1/2}$ parameter plane 
of the mSUGRA model, with
$\tan\beta =45$, $A_0=0$ and $\mu <0$.
The red (magenta) region is
excluded by theoretical (experimental) constraints. The region below
the magenta contour has $m_h <114.1$ GeV, in violation of Higgs mass
limits from LEP2.
}
\label{fig:45m}}

Finally, in Fig.~\ref{fig:52p}, we show the mSUGRA plane for
$\tan\beta =52$, $A_0=0$ and $\mu >0$. Again, the reach is diminished 
for low $m_0$, but remains substantial for large $m_0$, 
especially in the HB/FP region. As with the previous figures, the 
reach extends to $m_{1/2}\sim 270$ GeV, corrresponding to
a value of $m_{\tg}\sim 750$ GeV.

\FIGURE[t]{\epsfig{file=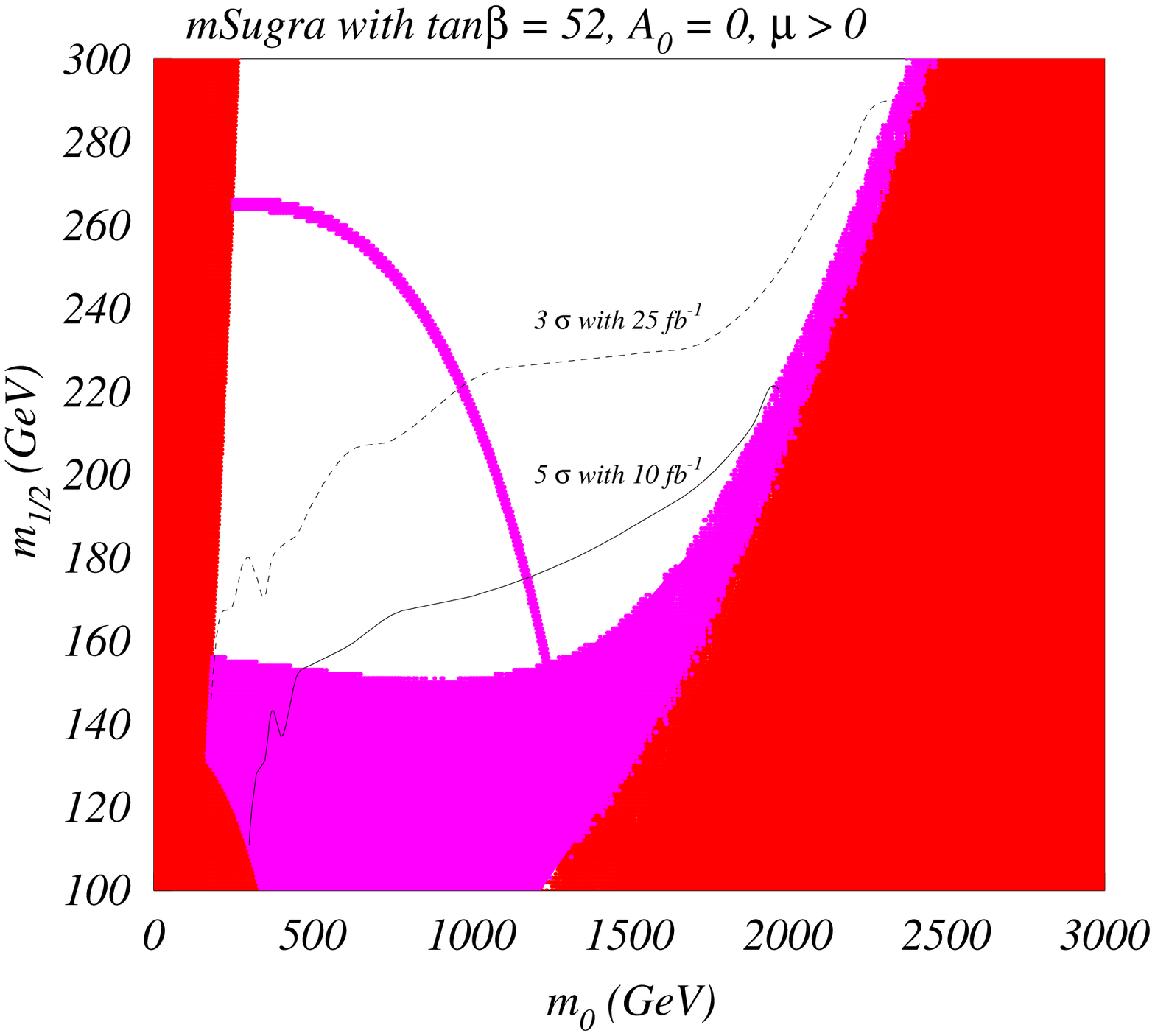,width=15cm} 
\caption{The reach of Fermilab Tevatron in the 
$m_0\ vs.\ m_{1/2}$ parameter plane 
of the mSUGRA model, with
$\tan\beta =52$, $A_0=0$ and $\mu >0$.
The red (magenta) region is
excluded by theoretical (experimental) constraints. The region below
the magenta contour has $m_h <114.1$ GeV, in violation of Higgs mass
limits from LEP2.
}
\label{fig:52p}}

{\it Summary:} In summary, we have evaluated the reach of the
Fermilab Tevatron collider for supersymmetry in the framework of the mSUGRA
model. The best signature for SUSY appears to be trilepton
events orginating from chargino/neutralino production, with subsequent 
leptonic decays. We have extended previous analyses into the large
$m_0$ region, where significant regions of parameter
space are accessible to Tevatron search experiments. This region
includes the intriguing HB/FP region, where squarks and sleptons
are heavy (thus ameliorating the SUSY flavor and CP problems),
while possibly maintaining naturalness\cite{fmm}.
In this region, since $\mu$ is decreasing, sparticle production
cross sections increase, and 
Tevatron experiments may be able to find evidence for SUSY out to
$m_{1/2}$ values as high as 200-280 GeV depending on 
the ultimate integrated luminosity which is achieved.

\section*{Acknowledgments}
 
We thank K. Matchev for discussions.
This research was supported in part by the U.S. Department of Energy
under contracts number DE-FG02-97ER41022 and DE-FG03-94ER40833.
	
%

\end{document}